\documentclass[11pt,a4paper]{article}
\usepackage{latexsym,amsthm,amssymb,amsmath}
\newtheorem{theorem}{Theorem}

\newtheorem{lemma}[theorem]{Lemma}
\newcommand{\be}{\begin{equation}}
\newcommand{\ee}{\end{equation}}
\newcommand{\bea}{\begin{eqnarray}}
\newcommand{\eea}{\end{eqnarray}}
\newcommand{\ba}{\begin{array}}
\newcommand{\ea}{\end{array}}
\newcommand{\bean}{\begin{eqnarray*}}
\newcommand{\eean}{\end{eqnarray*}}

\newcommand{\ep}{\epsilon}

\newcommand{\ga}{\gamma}

\newcommand{\la}{\lambda}
\newcommand{\om}{\omega}
\newcommand{\Om}{\Omega}
\newcommand{\de}{\delta}

\newcommand{\pa}{\partial}

\newcommand{\blm}{\begin{lemma}}
\newcommand{\elm}{\end{lemma}}
\newcommand{\bthm}{\begin{theorem}}
\newcommand{\ethm}{\end{theorem}}

\begin{document}

\title
 {\sc Kernel Formula Approach to the Universal Whitham Hierarchy
\/}
\author
{\sc   Hsin-Fu Shen$^1$, Niann-Chern Lee$^2$, and Ming-Hsien Tu$^3$
\footnote{phymhtu@ccu.edu.tw} \\
    $^1$
  {\it Department of Mechanical Engineering, WuFeng Institute of Technology,\/}\\
  {\it Chiayi 621, Taiwan\/},\\
  $^2$
  {\it General Education Center, National Chinyi University of Technology\/},\\
  {\it Taichung 411, Taiwan\/}\\
    $^3$
  {\it Department of Physics, National Chung Cheng University\/},\\
   {\it Chiayi 621, Taiwan\/}\/}
\date{\today}
\maketitle
\begin{abstract}
We derive the dispersionless Hirota equations of the universal Whitham hierarchy from the
kernel formula approach  proposed by Carroll and Kodama. Besides, we also verify the
associativity equations in this hierarchy from the dispersionless Hirota equations
and give a realization of the associative algebra with structure constants expressed
in terms of the residue formulas.

\end{abstract}

\newpage
\section{ Introduction}
Dispersionless integrable hierarchies have been an active subject of research in recent years.
The dispersionless KP (dKP) hierarchy\cite{Kod88,KG89,TT92,TT95}  and dispersionless
Toda (dToda) hierarchy\cite{TT91,TT93} are now the prototype systems in several branches of physics and mathematics,
such as topological field theories, matrix models, string theory, Laplacian growth problem and conformal
maps (see, e.g., \cite{AK96,BK00,BR03,BS03,Kr92,Kr94,WWZ05,WZ00,Za01,Za07}, and references therein).
The solution of the dispersionless integrable hierarchies
can be characterized by a $\tau$-function whose logarithm, $F=\log\tau$, called free energy
satisfies a set of dispersionless Hirota (dHirota) equations and gives the solutions to
the associativity equations or Witten-Dijkgraaf-Verlinde-Verlinde (WDVV) equations introduced
in the context of topological field theories\cite{Du96,DVV91,Wi90}.

It was shown\cite{Boy01} that the associativity equations are encoded in the dHirota equations
for the dKP and the dToda hierarchies. There are several works devoted to derive the
corresponding dHirota equations.
For the dKP hierarchy, Takasaki and Takebe\cite{TT95} derived them by taking the dispersionless limit of the
differential Fay identity. Later, Carroll and Kodama\cite{CK95} studied it from the approach of kernel formulas.
As for the dToda hierarchy, Wiegmann and Zabrodin et al.\cite{WZ00} investigated the dHirota equations in the language
of conformal mappings. Teo\cite{Teo03} derived the dHirota equations for both the dKP and dToda hierarchies
in the content of complex analysis using the notion of Faber polynomials and Grunsky coefficients.

In this work we shall focus on the universal Whitham hierarchy which was introduced by Krichever\cite{Kr94} as a
universal framework for both dispersionless integrable systems and Whitham modulation equations.
In particular, the genus-zero case is the main target of recent researches
(see e.g., \cite{AMM06,AM06,GMM03,KWWA04,MMA06,TT07,TT08}) since the universal Whitham hierarchy is
the master equation of many other dispersionless integrable systems. Based on the work of
Carroll and Kodama\cite{CK95} (see also \cite{CT06}) we like to generalize the kernel
formula approach for the dKP hierarchy to derive the dHirota equations of the
 universal Whitham hierarchy. Moreover, motivated by the work of Boyarsky et al.\cite{Boy01},
 we intend to provide a direct verification of associative equations in the universal
 Whitham hierarchy from the dHirota equations point of view.

This paper is organized as follows.
In the next section , we briefly recall the Lax formulation for the universal Whitham hierarchy
of zero genus  case. In section 3, the kernel formula approach is employed to derive the
dHirota equations of the universal Whitham hierarchy.
In Section 4, following the work of \cite{Boy01} , we will prove
the associativity equation of the universal Whitham hierarchy from these dHirota equations
and give a realization of the associative algebra.
The situations of the finite-dimensional reduction are also briefly discussed.
Section 5 is devoted to the concluding remarks.

\section{ Lax Formulation of the Universal Whitham Hierarchy }
The universal Whitham hierarchy of genus zero with $N+1$ marked points at
$q_\alpha (\alpha=1,\cdots,N)$ and $\infty$ on Riemann sphere is defined by
the Lax functions\cite{Kr94}
\bean
\la_0(p)&=&p+\sum_{j=2}u_{j}p^{-j+1}\\
\la_\alpha(p)&=&\frac{r_\alpha}{p-q_\alpha}+\sum_{j=1}u_{\alpha j}(p-q_\alpha)^{j-1},\quad
\alpha=1,\cdots,N
\eean
where $u_j$ and $u_{\alpha j}$ are functions of the time variables
$t_{0n} (n=1,2,\cdots)$ and $t_{\alpha n} (\alpha=1,\cdots,N; n=0,1,\cdots)$.
One can regard $\la_0$ as a map defined in domain $\Om_0$ containing $p=\infty$,
while $\la_\alpha$ in $\Om_\alpha$ containing $p=q_\alpha$. $\la_0$ and $\la_\alpha$
satisfy the Lax equations
\be
\pa_{\alpha n}\la_\beta(p)=\{B_{\alpha n}(p), \la_\beta(p)\},\quad
\pa_{\alpha n}=\pa/\pa t_{\alpha n}
\ee
where the Poisson bracket is defined by $\{f,g\}=\pa_pf\pa_{01}g-\pa_{01}f\pa_pg$ and
\bean
B_{0n}(p)&=&(\la_0^n)_{(0,\geq 0)}\\
B_{\alpha n}(p)&=&(\la_\alpha^n)_{(\alpha, <0)},\quad n=1,2,\cdots\\
B_{\alpha 0}(p)&=&-\log(p-q_\alpha).
\eean
Following the notation used in \cite{AMM06,TT07} we shall denote $()_{(0,\geq 0)}$ and $()_{(\alpha,<0)}$ as
the projections to nonnegative powers in $p$ and negative powers in $(p-q_\alpha)$, respectively.
Therefore
\bean
\la_0^n(p)&=&B_{0n}(p)+O(p^{-1}),\quad p\to \infty,\\
\la_\alpha^n(p)&=&B_{\alpha n}(p)+O(1),\quad p\to q_\alpha.
\eean
In particular, $B_{01}(p)=p$ and $B_{\alpha 1}(p)=r_\alpha/(p-q_\alpha)$.
The Lax equations can be written as the Zakharov-Shabat equations
\[
\pa_{\beta n}B_{\alpha m}(p)-\pa_{\alpha m}B_{\beta n}(p)+\{B_{\alpha m}(p), B_{\beta n}(p)\}=0
\]
which is equivalent to $\om\wedge \om=0$ with
\[
\om=\sum_{n=1}^\infty dB_{0n}\wedge dt_{0n}+
\sum_{\alpha=1}^N\sum_{n=0}^\infty dB_{\alpha n}\wedge dt_{\alpha n}.
\]
Introducing Darboux coordinate $(\la_0,M_0)$ in $\Om_0$ and
$(\la_\alpha,M_\alpha)$ in $\Om_\alpha$, we have
\bean
\om&=&d\la_0\wedge dM_0\qquad \mbox{in}\quad \Om_0\\
&=&d\la_\alpha\wedge dM_\alpha\qquad \mbox{in}\quad \Om_\alpha
\eean
where the Orlov-Schulman functions $M_0$ and $M_\alpha$ are defined by\cite{AMM06,TT07}
\bean
M_0&=&\sum_{n=1}nt_{0n}\la_0^{n-1}+\frac{t_{00}}{\la_0}+\sum_{n=1}\la_0^{-n-1}v_{0n},\quad
t_{00}=-\sum_{\alpha=1}^Nt_{\alpha 0} \\
M_\alpha&=&\sum_{n=1}nt_{\alpha n}\la_\alpha^{n-1}+\frac{t_{\alpha 0}}{\la_\alpha}+
\sum_{n=1}\la_\alpha^{-n-1}v_{\alpha n},
\eean
satisfies
\[
\pa_{\alpha n}M_\beta=\{B_{\alpha n}, M_\beta\}
\]
and the canonical relation $\{\la_0,M_0\}=\{\la_\alpha,M_\alpha\}=1$.

From the expressions of the symplectic two-from $\om$ we have, in $\Om_0$,
\[
d\left(M_0d\la_0+\sum_{n=1}^\infty B_{0n} dt_{0n}+
\sum_{\alpha=1}^N\sum_{n=0}^\infty B_{\alpha n} dt_{\alpha n}\right)=0.
\]
Therefore, there exists a $S$-function, $S_0$, such that
\[
dS_0=M_0d\la_0+\sum_{n=1}^\infty B_{0n} dt_{0n}+
\sum_{\alpha=1}^N\sum_{n=0}^\infty B_{\alpha n} dt_{\alpha n}.
\]
On the other hand, in $\Om_\alpha$, we have
\[
dS_\alpha=M_\alpha d\la_\alpha+\sum_{n=1}^\infty B_{0n} dt_{0n}+
\sum_{\alpha=1}^N\sum_{n=0}^\infty B_{\alpha n} dt_{\alpha n}.
\]
It turns out that the Orlov-Schulman functions $M_0$ and $M_\alpha$ have the expressions
\[
M_0=\frac{\pa S_0(t)}{\pa \la_0},\quad M_\alpha=\frac{\pa S_\alpha(t)}{\pa \la_\alpha}
\]
then the $S$-functions admit expansions of the form
\bean
S_0&=&\sum_{n=1}t_{0n}\la_0^n+t_{00}\log \la_0-\sum_{n=1}\frac{\la_0^{-n}}{n}v_{0n},\\
S_\alpha&=&\sum_{n=1}t_{\alpha n}\la_\alpha^n+t_{\alpha 0}\log \la_\alpha+\phi_\alpha
-\sum_{n=1}\frac{\la_\alpha^{-n}}{n}v_{\alpha n}.
\eean

The Hamilton-Jacobi equations are defined by
\[
\pa_{\alpha n}S_\beta=B_{\alpha n}(p_\beta)=B_{\alpha n}(\pa_{01}S_\beta)
\]
hence
\bean
B_{0n}&=&\pa_{0n} S_0(t)=
\la_0^n-\sum_{m=1}\frac{\la_0^{-m}}{m}\pa_{0n}v_{0m}\\
&=&\pa_{0n}S_\alpha(t)=
\pa_{0n}\phi_\alpha-\sum_{m=1}\frac{\la_\alpha^{-m}}{m}\pa_{0n}v_{\alpha m}\\
B_{\alpha n}&=&\pa_{\alpha n} S_0=
-\sum_{m=1}\frac{\la_0^{-m}}{m}\pa_{\alpha n}v_{0m}\\
&=&\pa_{\alpha n} S_\beta=
\de_{\alpha\beta}\la_\beta^{n}+\pa_{\alpha n}\phi_\beta
-\sum_{m=1}\frac{\la_\beta^{-m}}{m}\pa_{\alpha n}v_{\beta m}\\
B_{\alpha 0}&=&\pa_{\alpha 0} S_0=
-\log\la_0-\sum_{m=1}\frac{\la_0^{-m}}{m}\pa_{\alpha 0}v_{0 m}\\
&=&\pa_{\alpha 0}S_\beta=
\de_{\alpha\beta}\log\la_\beta+\pa_{\alpha 0}\phi_\beta
-\sum_{m=1}\frac{\la_\beta^{-m}}{m}\pa_{\alpha 0}v_{\beta m}.
\eean
The above equations provide some relations between the coefficients
 $(u_j,u_{j\alpha},r_\alpha,q_\alpha)$ and $(\phi_\alpha,v_{\alpha n})$.
For instance, from $B_{01}=p$ we have
\bean
p=\pa_{01}S_0(\la)=\la_0-\sum_{m=1}\frac{\la_0^{-m}}{m}\pa_{01}v_{0m},\qquad \mbox{in}\quad \Om_0\\
p=\pa_{01}S_\alpha(\la)=\pa_{01}\phi_\alpha-\sum_{m=1}\frac{\la_\alpha^{-m}}{m}\pa_{01}v_{\alpha m}
,\qquad \mbox{in}\quad \Om_\alpha
\eean
which implies $u_2=\pa_{01}v_{01}$, $q_\alpha=\pa_{01}\phi_\alpha$
and $r_\alpha=-\pa_{01}v_{\alpha 1}$. On the other hand,
$B_{\alpha 0}(p_\alpha)=\pa_{\alpha 0}S_\alpha=-\log(p_\alpha-q_\alpha)$ which yields
$r_\alpha=e^{-\pa_{\alpha 0}\phi_\alpha}$. In fact, there exist a $F$ function
\cite{Kr94,AMM06,TT07}
such that
\be
v_{0n}=F_{0n},\quad v_{\alpha n}=F_{\alpha n},\quad
\phi_\alpha=-F_{\alpha 0}+\sum_{\beta=1}^\alpha t_{\beta 0}\log(-1),
\label{def-F}
\ee
where $F_{\alpha n}=\pa_{\alpha n}F$. Then we see that $u_2=\pa^2_{01}F$, the dKP potential.
And from $r_\alpha=e^{-\pa_{\alpha 0}\phi_\alpha}$, $\pa_{01}\pa_{\alpha 1}F-e^{\pa^2_{\alpha 0}F}=0$,
which is just the dToda field equation.
We remark here that the Lax equation for the $t_{\alpha 0}$ flow is $\pa_{\alpha 0} \la = \{B_{\alpha 0},\la \}=
\frac{-1}{p}\frac{\pa \la}{\pa t_{01}}$ if setting $q_\alpha$ to be zero.
The Poisson bracket may be written as, not so strictly, $\{f,g\}=-(p\frac{\pa f}{\pa p}\frac{\pa g}{\pa t_{\alpha 0}}
-p\frac{\pa f}{\pa t_{\alpha 0}}\frac{\pa g}{\pa p})$. The time variable $-t_{\alpha 0}$ seems to play the role of the
parameter $s$ in the usual Poisson bracket  of the dToda hierarchy\cite{TT91,TT93,TT95}.
Therefore the universal Whitham hierarchy can be viewed as an assembly
of one dKP hierarchy and $N$ copies of dToda-like hierarchies.

\section{Carroll-Kodama Approach }
\subsection{Kernel Formulas}
In contrast to the dKP and dToda hierarchy, the Whitham hierarchy contains
$N+1$ Lax functions $\la _0$ and $\la _\alpha$ that are defined in domains $\Om_0$ and $\Om_\alpha (\alpha=1,\cdots,N)$,
respectively. To derive the kernel formulas of the universal Whitham hierarchy we
shall generalize the method of kernel formula to the domains $\Om_0$, $\Om_\alpha$,
$\Om_0\cap \Om_\alpha$ and $\Om_\alpha\cap \Om_\beta (\alpha\neq \beta)$.

In $\Om_0$, from the expression of $B_{01}=p=\pa_{01}S_0$, we have
\[
\la_0-p-\sum_{m=1}\frac{\la_0^{-m}}{m}F_{01,0m}=0.
\]
Multiplying above by $\la_0^{n-1}\pa_p\la_0$ and taking the projection $()_{(0,\geq 0)}$
then we obtain
\be
\pa_pQ_{0,n+1}(\la)-p\pa_pQ_{0,n}(\la)-\sum_{m=1}^{n-1}\frac{F_{01,0m}}{m}\pa_pQ_{0,n-m}(\la)=0,
\quad n\geq 1
\label{rec-1}
\ee
where $Q_{0,n}(\la)=B_{0n}(\la)/n$. Taking the summation $\sum_{n=0}\mu^{-n}$ and noting that
$\pa_p Q_{0,1}=1$ we have
\[
\left(\mu-p_0(\la)-\sum_{m=1}^\infty\frac{F_{01,0m}}{m}\mu^{-m}\right)\sum_{n=1}\pa_pQ_{0,n}(\la)\mu^{-n}=1
\]
or
\be
\frac{1}{p_0(\mu)-p_0(\la)}=\sum_{n=1}\pa_pQ_{0,n}(\la)\mu^{-n}.
\label{g1}
\ee
Integrating above with respect to $p_0(\la)$ by fixing the
integration constant at $\mu\to\infty$, we find
\be
\log\frac{\mu}{p_0(\mu)-p_0(\la)}=\sum_{n=1}\frac{B_{0,n}(\la)}{n}\mu^{-n}
\label{ker1}
\ee
Rewriting the expressions of $p_0=\pa_{01}S_0$ and $B_{0n}=\pa_{0n}S_0$
in kernel formula (\ref{ker1}) as
\bean
p_0(\la)&=&\la-D_0(\la)F_{01},\\
B_{0n}(\la)&=&\la^n-D_0(\la)F_{0n}
\eean
where we have used the symbols $D_\alpha(z)=\sum_{n=1}^\infty\frac{z^{-n}}{n}\pa_{\alpha n}$.
Then the kernel formula (\ref{ker1}) becomes
\be
e^{D_0(\mu)D_0(\la)F}=1-\frac{\pa_{01}(D_0(\mu)-D_0(\la))F}{\mu-\la}.
\label{dH1}
\ee

In $\Om_\alpha$, from $B_{\alpha n}=\pa_{\alpha n}S_\alpha$, we have
\[
B_{\alpha 1}=\frac{r_\alpha}{p-q_\alpha}=\la_\alpha+\pa_{\alpha 1}\phi_\alpha
-\sum_{m=1}\frac{\la_\alpha^{-m}}{m}F_{\alpha 1, \alpha m}
\]
Multiplying above by $\la_\alpha^{n-1}\pa_p\la_\alpha$ and taking the projection $()_{(0,\leq -2)}$
then we obtain
\be
\pa_pQ_{\alpha,n+1}(\la)-\left(\frac{r_\alpha}{p-q_\alpha}-\pa_{\alpha 1}\phi_\alpha\right)\pa_pQ_{\alpha,n}(\la)
-\sum_{m=1}^{n-1}\frac{F_{\alpha 1,\alpha m}}{m}\pa_pQ_{\alpha,n-m}(\la)=0,
\quad n\geq 1
\label{rec-2}
\ee
where $Q_{\alpha,n}(\la)=B_{\alpha n}(\la)/n$.
 Taking the summation $\sum_{n=0}\mu^{-n}$ and noting that
$\pa_p Q_{\alpha,1}=-\frac{r_\alpha}{(p-q_\alpha)^2}$ we have
\[
\left(\mu-\left(\frac{r_\alpha}{p-q_\alpha}-\pa_{\alpha 1}\phi_\alpha\right)
-\sum_{m=1}^\infty\frac{F_{\alpha 1,\alpha m}}{m}\mu^{-m}\right)
\sum_{n=1}\pa_pQ_{\alpha,n}(\la)\mu^{-n}=-\frac{r_\alpha}{(p-q_\alpha)^2}
\]
or
\be
\frac{p_\alpha(\mu)-q_\alpha}{(p_\alpha(\mu)-p_\alpha(\la))(p_\alpha(\la)-q_\alpha)}
=\sum_{n=1}\pa_pQ_{\alpha,n}(\la)\mu^{-n}.
\label{g2}
\ee
Integrating above with respect to $p_\alpha(\la)$ by fixing the
integration constant at $\mu\to\infty$, we find
\be
\log\frac{p_\alpha(\la)-q_\alpha}{p_\alpha(\la)-p_\alpha(\mu)}
=\sum_{n=1}\frac{B_{\alpha,n}(\la)}{n}\mu^{-n}
\label{ker2}
\ee
Rewriting the expressions of $p_\alpha-q_\alpha$, $p_\alpha$
and $B_{\alpha n}$ in kernel formula (\ref{ker2}) as
\bean
p_\alpha-q_\alpha&=&e^{-\pa_{\alpha 0}S_\alpha(\la)}=
-\frac{1}{\la}e^{\pa_{\alpha 0}(\pa_{\alpha 0}+D_\alpha(\la))F}\\
p_\alpha(\la)&=&\pa_{01}S_\alpha(\la)=\pa_{01}\phi_\alpha-D_\alpha(\la)F_{01}\\
B_{\alpha n}&=&\pa_{\alpha n}S_\alpha(\la)=
\la^n+\pa_{\alpha n}\phi_\alpha-D_{\alpha}(\la)F_{\alpha n}
\eean
then the kernel formula (\ref{ker2}) becomes
\be
e^{(\pa_{\alpha 0}+D_\alpha(\mu))(\pa_{\alpha 0}+D_\alpha(\la))F}=
-\frac{\mu\la}{\mu-\la}\pa_{01}(D_\alpha(\mu)-D_\alpha(\la))F.
\label{dH2}
\ee

In $\Om_0\cap \Om_\alpha$, let us rewrite (\ref{ker1}) as
\[
\frac{\mu}{\pa_{01}S_0(\mu)-\pa_{01}S_0(\la)}=e^{\sum_{n=1}\frac{\pa_{0n}S_0(\la)}{n}\mu^{-n}}.
\]
By replacing  $S_0(\la)$ by $S_\alpha(\la)$ and using the relations
\bean
\pa_{01}S_\alpha(\la)&=&\pa_{01}\phi_\alpha-D_\alpha(\la)F_{01}\\
\pa_{0n}S_\alpha(\la)&=&\pa_{0n}\phi_\alpha-D_\alpha(\la)F_{0n}
\eean
then we have
\be
\mu e^{D_0(\mu)(\pa_{\alpha 0}+D_\alpha(\la))F}=
\mu-\pa_{01}(D_0(\mu)-\pa_{\alpha 0}-D_\alpha(\la))F.
\label{dH3}
\ee
On the other hand, one can derive (\ref{dH3}) from the kernel formula (\ref{ker2}) by
expressing it as
\[
\frac{e^{-\pa_{\alpha 0}S_\alpha(\la)}}{\pa_{01}S_\alpha(\la)-\pa_{01}S_\alpha(\mu)}=
e^{\sum_{n=1}\frac{\pa_{\alpha n}S_\alpha(\la)}{n}\mu^{-n}}.
\]
After replacing $S_\alpha(\la)$ by $S_0(\la)$, then the above equation becomes
\[
\la e^{D_0(\la)(\pa_{\alpha 0}+D_\alpha(\mu))F}=\la-\pa_{01}(D_0(\la)-\pa_{\alpha 0}-D_\alpha(\mu))F
\]
which, after making an exchange $\la\leftrightarrow\mu$, coincides with (\ref{dH3}).

In $\Om_\alpha\cap \Om_\beta$ with $\alpha\neq \beta$,
we rewrite (\ref{ker2}) as
\[
\frac{e^{-\pa_{\alpha 0}S_\alpha(\la)}}{\pa_{01}S_\alpha(\la)-\pa_{01}S_\alpha(\mu)}=
e^{\sum_{n=1}\frac{\pa_{\alpha n}S_\alpha(\la)}{n}\mu^{-n}}.
\]
By replacing  $S_\alpha(\la)$ by $S_\beta(\la)$,
then above equation becomes
\be
\frac{e^{-\pa_{\alpha 0}S_\beta(\la)}}{\pa_{01}S_\beta(\la)-\pa_{01}S_\alpha(\mu)}=
e^{\sum_{n=1}\frac{\pa_{\alpha n}S_\beta(\la)}{n}\mu^{-n}}.
\label{ker4}
\ee
Since
\bean
e^{-\pa_{\alpha 0}S_\beta(\la)}&=&e^{-\pa_{\alpha 0}\phi_\beta+\pa_{\alpha 0}D_\beta(\la)F}=
-\ep_{\alpha\beta}e^{\pa_{\alpha 0}(\pa_{\beta 0}+D_\beta(\la))F},\\
\pa_{01}S_\beta(\la)-\pa_{01}S_\alpha(\mu)&=&\pa_{01}(D_\alpha(\mu)+\pa_{\alpha 0}-\pa_{\beta 0}-D_\beta(\la))F,\\
\pa_{\alpha n}S_\beta(\la)&=&\pa_{\alpha n}\phi_\beta-
\sum_{m=1}\frac{F_{\alpha n,\beta m}}{m}\la^{-m},
\eean
where $\phi_\beta$ is defined by (\ref{def-F}) and
\[
\ep_{\alpha\beta}=\left\{
\ba{c}
+1,\quad \alpha\leq \beta\\
-1,\quad \alpha>\beta
\ea
\right.
\]
 Thus eq.(\ref{ker4}) can be expressed as
\be
\ep_{\alpha\beta}e^{(\pa_{\alpha 0}+D_\alpha(\mu))(\pa_{\beta 0}+D_\beta(\la))F}
=-\pa_{01}(D_\alpha(\mu)+\pa_{\alpha 0}-\pa_{\beta 0}-D_\beta(\la))F
\label{dH4}
\ee
Eqs. (\ref{dH1}), (\ref{dH2}), (\ref{dH3}) and (\ref{dH4}) constitute the dispersionless
Hirota equations of the genus-zero universal Whitham hierarchy.
They describe algebraic relations between second derivatives of the
free energy $F$. Remarkably, our results coincide with that by Takasaki and Takebe \cite{TT07}
who derived these equations from the Hamilton-Jacobian equations in the context of the Faber polynomials.
\subsection{Dispersionless Hirota Equations}
Using the parametrization\cite{TT08}:
$\hat{t}_{0n}=nt_{0n}$, $\hat{t}_{\alpha 0}=t_{\alpha 0}$ and $\hat{t}_{\alpha n}=nt_{\alpha n}$.
the corresponding time derivative is written as $\hat{\pa}_{\alpha n}=\frac{\pa}{\pa \hat{t}_{\alpha n}}$.
Then the four dispersionless Hirota equations can be rewritten as\cite{TT07} :
\be
\begin{split}
\hat{D}_0(\mu)\hat{D}_0(\la)F&=\log \frac{p_0(\mu)-p_0(\la)}{\mu-\la},\\
\hat{D}_{\alpha}(\mu)\hat{D}_{\alpha}(\la)F&=\log \frac{\mu \la (p_{\alpha}(\mu)-p_{\alpha}(\la))}{\mu-\la},\\
\hat{D}_0(\mu)\hat{D}_{\alpha}(\la)F&=\log \frac{p_0(\mu)-p_{\alpha}(\la)}{\mu}, \\
\hat{D}_{\alpha}(\mu)\hat{D}_{\beta}(\la)F&=\log\frac{p_{\alpha}(\mu)-p_{\beta}(\la)}{\varepsilon_{\alpha \beta}},
\label{dd4}
\end{split}
\ee
where we have use the new symbols :
$\hat{D}_0(z)=\sum_{n=1}^\infty\frac{z^{-n}}{n}\pa_{0n}=\sum_{n=1}^\infty z^{-n}\hat{\pa}_{0n}$, and
$\hat{D}_\alpha(z)=\pa_{\alpha 0}+\sum_{n=1}^\infty\frac{z^{-n}}{n}\pa_{\alpha n}=
\sum_{n=0}^\infty z^{-n}\hat{\pa}_{\alpha n}$.
This set of equations will be our main interests in the following discussion.
Note that they have the same form if applying the differentiation operator
$\hat{D}_{\de}(\nu)$ on them. The result can be summarized as a single equation:
\be
\hat{D}_{\alpha}(\mu)\hat{D}_{\beta}(\la)\hat{D}_{\de}(\nu)F=
-\frac{(\hat{D}_{\alpha}(\mu)-\hat{D}_{\beta}(\la))\hat{D}_{\de}(\nu)\hat{F}_{01}}
{p_{\alpha}(\mu)-p_{\beta}(\la)},\\
\label{ddd4}
\ee
where $\alpha, \beta, \de = 0,1,2,\cdots,N$. In fact, we may rewrite (\ref{dd4}) as a set
of equations satisfied by the second derivatives of $F$.
Letting $\mu\to \lambda$ in (\ref{dd4}) we have
\bean
k\hat{F}_{0k,01}&=&P_{k+1}(X),\quad X_1=0, X_{j\geq 2}=\sum_{n+m=j}\hat{F}_{0n,0m}\\
k\hat{F}_{\alpha k,01}&=&e^{\hat{F}_{0\alpha 0,\alpha 0}}P_{k-1}(X),
\quad X_{j\geq 1}=\sum_{n+m=j}\hat{F}_{\alpha n,\alpha m}\\
k\hat{F}_{0k,01}&=&P_{k+1}(X)-\sum_{l=1}^{k+1}Y_lP_{k+1-l}(X),
\quad X_{j\geq 1}=\sum_{n+m=j}\hat{F}_{0 n,\alpha m},
Y_j=\sum_{n+m=j}n\hat{F}_{0 n,\alpha m}\\
k\hat{F}_{\alpha k,01}&=&\mp e^{\hat{F}_{\alpha 0,\beta 0}}\sum_{l=1}^kY_lP_{k-l}(X),
\quad X_{j\geq 1}=\sum_{n+m=j}\hat{F}_{\alpha n,\beta m},
Y_j=\sum_{n+m=j,n\geq 1}n\hat{F}_{\alpha n,\beta m}
\eean
where $k\geq 1$ and $P_k(X)$ are Schur polynomials defined by
$e^{\sum_{j=1}X_js^j}=\sum_{k=0}P_k(X)s^k$. The first nontrivial one of each dHirota equation is
 given below
\bean
\hat{F}_{03,01}-\hat{F}_{02,02}-\hat{F}^2_{01,01}/2&=&0\\
\hat{F}_{\alpha 1,01}-e^{\hat{F}_{\alpha 0,\alpha 0}}&=&0\\
\hat{F}_{01,01}+\hat{F}_{02,\alpha 0}+\hat{F}^2_{01,\alpha 0}/2&=&0\\
\hat{F}_{\alpha 1,01}\pm e^{\hat{F}_{\alpha 0,\beta 0}}\hat{F}_{\alpha 1,\beta 0}&=&0
\eean
We remark that, in terms of $t_{\alpha n}$, the first and the second equations are
 just the dKP and dToda equations, respectively.

\section{ Associativity Equations in the Whitham Hierarchy}

\subsection{WDVV equations}
Let $F$ be a function of the time variables $\hat{t}_J$ where the subscribes $J=\{(0,n),(\alpha,m)\}$
with $\alpha=1,\cdots,N; n\geq 1,m\geq 0$
and denoting $\hat{F}_{J}=\hat{\pa}_JF$, $\hat{F}_{KL}=\hat{\pa}_J\hat{\pa}_KF$, etc. Let us define
$\hat{t}_I=\hat{t}_{01}$ and
introduce the nondegenerate metric $\hat{\eta}_{JK}=\hat{F}_{JKI}$ which provides a transformation between
$\{ \hat{t}_J\}$ and $\{\hat{F}_{KI}\}$. Thus
\be
\hat{F}_{JKL}=\sum_A\frac{\pa \hat{F}_{JK}}{\pa \hat{F}_{AI}}\frac{\pa \hat{F}_{AI}}{\pa \hat{t}_L}
=\sum_A\hat{C}^A_{JK}\hat{\eta}_{AL}
\label{FIJK}
\ee
where the coefficients $\hat{C}_{JK}^A$ defined above connect $\hat{F}_{JKL}$ and $\hat{\eta}_{AL}$.
Notice that setting $J=I$ one has $\hat{\eta}_{KL}=\sum_A\hat{C}^A_{IK}\hat{\eta}_{AL}$
and hence $\hat{C}_{IK}^A=\delta_K^A$.
If $\hat{C}_{JK}^L$ are structure constants of an associative algebra generated by  $\{\hat{\phi}_J\}$, i.e.,
$\hat{\phi}_J\cdot\hat{\phi}_K=\sum_L\hat{C}_{JK}^L\hat{\phi}_L$, then the associativity condition
$(\hat{\phi}_J\cdot\hat{\phi}_K)\cdot\hat{\phi}_L=\hat{\phi}_J\cdot(\hat{\phi}_K\cdot\hat{\phi}_L)$ yields
$\hat{C}_{JK}^M\hat{C}_{ML}^N=\hat{C}_{JM}^N\hat{C}_{KL}^M$, or, equivalently,
\[
\sum_M\hat{C}_{JK}^M\hat{F}_{MLN}=\sum_M\hat{C}_{KL}^M\hat{F}_{JMN}.
\]
The above equation is known as the WDVV equations arising in the context of topological
field theory\cite{Du96,DVV91,Wi90}. Define a four indices quantity $\hat{X}_{JKMN}=\sum_L\hat{C}_{JK}^L\hat{F}_{LMN}$, then
the WDVV equations can be replaced by the condition that $\hat{X}_{JKMN}$ is symmetric
with respect to permutations of any indices. Using the operator
$\hat{D}_\alpha (z)$ introduced in the previous section,  it is convenient
to define the generating functions for $\hat{F}_{JK}$, $\hat{F}_{JKL}$,
$\hat{C}_{JK}^L$ and $\hat{X}_{JKLM}$ as follows:
\bea
\hat{D}_{\alpha}(u)\hat{D}_{\beta}(v)F&=&\sum_{n,m=0}{u^{-n}}{v^{-m}}\hat{F}_{\alpha n,\beta m},
\label{def2f}\\
\hat{D}_{\alpha}(u)\hat{D}_{\beta}(v)\hat{D}_{\ga}(w)F&=&\sum_{n,m,l=0}{u^{-n}}{v^{-m}}
{w^{-l}}\hat{F}_{\alpha n,\beta m,\gamma l},\label{def3f}\\
\hat{C}^J(u_\alpha,v_\beta)&=&\sum_{n,m=0}{u^{-n}}{v^{-m}}\hat{C}_{\alpha n,\beta m}^J,\label{defc}\\
\hat{X}(u_\alpha,v_\beta,w_\gamma,z_\delta)&=&\sum_{n,m,k,l=0}
{u^{-n}}{v^{-m}}{w^{-k}}{z^{-l}}\hat{X}_{\alpha n,\beta m,\gamma k,\delta l}.\label{defx}
\eea
Hence the WDVV equations can be expressed in terms of the generating function as
\[
\hat{X}(u_\alpha,v_\beta,w_\gamma,z_\delta)=\hat{X}(u_\alpha,w_\gamma,v_\beta,z_\delta).
\]

\subsection{From dHirota to WDVV}

Applying $\hat{\pa}_{\alpha l}$ on the first line of dHirota equation(\ref{dd4}), we get
\[
\hat{D}_0(u)\hat{D}_0(v)\hat{\pa}_{\alpha l}F=-\frac{(\hat{D}_0(u)-\hat{D}_0(v))\hat{\pa}_{\alpha l}\hat{F}_{01}}
{p_0(u)-p_0(v)}
\]
The left hand side is
\bean
\hat{D}_0(u)\hat{D}_0(v)\hat{\pa}_{\alpha l}F&=&\sum_{n,m=1}{u^{-n}}{v^{-m}}\hat{F}_{0n,0m,\alpha l}\\
&=&\sum_J\sum_{n,m=1}{u^{-n}}{v^{-m}}\hat{C}_{0n,0m}^J\hat{F}_{J,\alpha l,I}\\
&=&\sum_J\hat{C}^J(u_0,v_0)\hat{F}_{J,\alpha l,I},
\eean
while the right hand side is
\[
-\frac{(\hat{D}_0(u)-\hat{D}_0(v))\hat{\pa}_{\alpha l}\hat{F}_{01}}{p_0(u)-p_0(v)}=
-\sum_{k=1}\frac{u^{-k}-v^{-k}}{p_0(u)-p_0(v)}\hat{F}_{0k,\alpha l,I}.
\]
Hence, we have
\be
\hat{C}^J(u_0,v_0)=
\left\{
\ba{ll}
-\frac{u^{-n}-v^{-n}}{p_0(u)-p_0(v)} & J=(0,n)\\
0 & J=\{(\alpha,n);\alpha\neq 0\}\\
\ea
\right.
\ee
By the same way, one can get the other structure constant generating functions from
the rest of equation(\ref{dd4}) . The formulas are listed below:
\[
\hat{C}^J(u_{\alpha},v_{\alpha})=
\left\{
\ba{ll}
-\frac{u^{-n}-v^{-n}}{p_{\alpha}(u)-p_{\alpha}(v)} & J=\{(\alpha,n);\alpha\neq 0\} \\
0 & J=(0,n),\\
\ea
\right.
\]
\[
\hat{C}^J(u_0,v_{\alpha})=
\left\{
\ba{ll}
\frac{-u^{-n}}{p_0(u)-p_{\alpha}(v)} & J=(0,n) \\
\frac{v^{-n}}{p_0(u)-p_{\alpha}(v)} & J=\{(\alpha,n);\alpha\neq 0\},\\
\ea
\right.
\]
and
\[
\hat{C}^J(u_\alpha,v_\beta)=
\left\{
\ba{ll}
\frac{-u^{-n}}{p_\alpha(u)-p_\beta(v)} & J=\{(\alpha,n);\alpha\neq 0\} \\
\frac{v^{-n}}{p_\alpha(u)-p_\beta(v)} & J=\{(\beta,n);\beta\neq 0\}.\\
\ea
\right.
\]
We remark here that by comparing above with (\ref{defc}) it is easy to show that
\bean
&&\hat{C}^{0k}_{0n,0m}=0,\quad k>n+m-1;
\qquad \hat{C}^{\alpha k}_{\alpha n,\alpha m}=0,\quad k>n+m+1\\
&&\hat{C}^{0k}_{0n,\alpha m}=0,\quad k>n-1;
\qquad \hat{C}^{\alpha k}_{0n,\alpha m}=0,\quad k>m\\
&&\hat{C}^{\alpha k}_{\alpha n,\beta m}=0,\quad k>n;
\qquad \hat{C}^{\beta k}_{\alpha n,\beta m}=0,\quad k>m
\eean
and hence the infinite sum in (\ref{FIJK}) involves only finite terms.

We can now show that any solution $F$ of the universal Whitham hierarchy obey the associativity equation with
the help of those structure constants. To show that the generating function (\ref{defx}) is totally symmetric
with respect to the permutation of $u_\alpha,v_\beta,w_\ga,z_\de$, however, it is enough to prove the symmetry
with respect to the permutation of $v_\beta$ and $w_\ga$ in (\ref{defx}), that is,
$\hat{X}(u_\alpha,v_\beta,w_\ga,z_\de)=\hat{X}(u_\alpha,w_\ga,v_\beta,z_\de)$.

For the case of $\alpha , \beta , \ga , \de \neq 0$, we need to prove that
\be
\begin{split}
&\quad (p_\alpha (u)-p_\ga(w))(\hat{D}_\alpha(u)-\hat{D}_\beta(v))\hat{D}_\ga(w)\hat{D}_\de(z)F\\
&= (p_\alpha (u)-p_\beta(v))(\hat{D}_\alpha(u)-\hat{D}_\ga(w))\hat{D}_\beta(v)\hat{D}_\de(z)F,
\end{split}
\ee
By virtue of (\ref{ddd4}), one can easily check that it is an identity.
Take as the second case : $\alpha = \beta = \ga = \de = 0$, we need to prove that
\begin{align*}
&\quad (p_0(u)-p_0(w))(\hat{D}_0(u)-\hat{D}_0(v))\hat{D}_0(w)\hat{D}_0(z)F\\
&= (p_0(u)-p_0(v))(\hat{D}_0(u)-\hat{D}_0(w))\hat{D}_0(v)\hat{D}_0(z)F,
\end{align*}
Consulting (\ref{ddd4}), one can see that it actually holds.
As for the case : $\alpha \neq \ga \neq \de $,$\beta =0$, the one needed to be proven is that
\begin{align*}
&\quad (p_\alpha (u)-p_\ga(w))(\hat{D}_0(v)-\hat{D}_\alpha(u))\hat{D}_\ga(w)\hat{D}_\de(z)F\\
&= (p_0(v)-p_\alpha(u))(\hat{D}_\alpha(u)-\hat{D}_\ga(w))\hat{D}_0(v)\hat{D}_\de(z)F
\end{align*}
One can also verify that it is true according to (\ref{ddd4}).
For the other cases , the calculation is similar to the above one, we omit that here.

\subsection{Realization of the Associative algebra}
To construct the generators $\{\hat{\phi}_J\}$ which realize the associative algebra with
structure constants $\hat{C}_{JK}^L$ obtained in the previous subsection we define
\be
\hat{\phi}_{\alpha n}(p)=
\left\{
\ba{ll}
\frac{d Q_{\alpha n}(p)}{dp},\quad & \alpha=0,1,\cdots,N; n\geq 1.\\
\frac{d B_{\alpha 0}(p)}{dp},\quad & \alpha=1,\cdots,N.\\
\ea
\right.
\ee
such that $\hat{\phi}_{0n}(p)$ are polynomials in $p$ and $\hat{\phi}_{\alpha n}(p)$
for $\alpha\neq 0$ are polynomials in $(p-q_\alpha)^{-1}$.
In particular, $\hat{\phi}_{01}(p)=1$, the unit element
of the algebra. Equations (\ref{rec-1}) and (\ref{rec-2}) provide the recursion relations
for $\hat{\phi}_{0n}(p)$ and $\hat{\phi}_{\alpha n}(p)$, respectively.
 There are four classes to be discussed.

In $\Omega_0$,from the kernel formula (\ref{g1}) we have
\be
\frac{1}{p_0(z)-p}=\sum_{n=1}{z^{-n}}\hat{\phi}_{0n}(p).
\label{p0}
\ee
Thus
\bean
\frac{1}{(p_0(z)-p)(p_0(w)-p)}&=&\sum_{n,m=1}{z^{-n}}{w^{-m}}\hat{\phi}_{0n}(p)\cdot \hat{\phi}_{0m}(p)\\
&=&-\frac{1}{p_0(z)-p_0(w)}\left(\frac{1}{p_0(z)-p}-\frac{1}{p_0(w)-p}\right)\\
&=&-\frac{1}{p_0(z)-p_0(w)}\left(\sum_{l=1}({z^{-l}-w^{-l}})\hat{\phi}_{0l}(p)\right)\\
&=&\sum_{l=1}\hat{C}^{0l}(z_0,w_0)\hat{\phi}_{0l}(p),
\eean
which together with (\ref{defc}) implies
\be
\hat{\phi}_{0n}\cdot\hat{\phi}_{0m}=\sum_{l=1}\hat{C}^{0l}_{0n,0m}\hat{\phi}_{0l}(p).
\ee
In $\Omega_{\alpha}$, note that the kernel formula (\ref{g2}) can be rewritten as the following form:
\be
\frac{1}{p_\alpha(w)-p}=
\sum_{m=0}{w^{-m}}\hat{\phi}_{\alpha m}(p).
\label{pa}
\ee
By the same way as above, we have
\[
\hat{\phi}_{\alpha n}\cdot\hat{\phi}_{\alpha m}=\sum_{l=0}\hat{C}^{\alpha l}_{\alpha n,\alpha m}\hat{\phi}_{\alpha l}(p).
\]
While in $\Omega_0 \cap \Omega_{\alpha}$, it's easy to see that
\bean
\frac{1}{p_0(z)-p}\frac{1}{p_\alpha(w)-p}
&=&\sum_{n=1,m=0}{z^{-n}}{w^{-m}}\hat{\phi}_{0n}(p)\cdot\hat{\phi}_{\alpha m}(p)\\
&=&-\frac{1}{p_0(z)-p_\alpha(w)}\left(\frac{1}{p_0(z)-p}-\frac{1}{p_\alpha(w)-p}\right)\\
&=&-\frac{1}{p_0(z)-p_\alpha(w)}\left(\sum_{n=1}z^{-n}\hat{\phi}_{0n}
-\sum_{n=0}w^{-n}\hat{\phi}_{\alpha n}\right)\\
&=&\sum_{n=1}\hat{C}^{0n}(z_0,w_{\alpha})\hat{\phi}_{0n}
+\sum_{n=0}\hat{C}^{\alpha n}(z_0,w_{\alpha})\hat{\phi}_{\alpha n},
\eean
which implies that
\[
\hat{\phi}_{0n}\cdot\hat{\phi}_{\alpha m}=\sum_{l=1}\hat{C}^{0l}_{0n,\alpha m}\hat{\phi}_{0l}(p)
+\sum_{l=0}\hat{C}^{\alpha l}_{0n,\alpha m}\hat{\phi}_{\alpha l}(p).
\]
Finally, in $\Omega_{\alpha} \cap \Omega_{\beta}$, we have
\[
\hat{\phi}_{\alpha n}\cdot\hat{\phi}_{\beta m}=\sum_{l=0}\hat{C}^{\alpha l}_{\alpha n,\beta m}\hat{\phi}_{\alpha l}(p)
+\sum_{l=0}\hat{C}^{\beta l}_{\alpha n,\beta m}\hat{\phi}_{\beta l}(p).
\]
Therefore, the generators $\{\hat{\phi}_J\}$ indeed satisfy the algebra
$\hat{\phi}_J\cdot \hat{\phi}_J=\sum_{L}\hat{C}^L_{JK}\hat{\phi}_L$ with structure
constants defined in the previous section.

Next let us derive the residue formulas for the third derivatives
of $F$ directly from the dHirota equations.
Applying $\hat{\pa}_{01}$ on the first dHirota equation(\ref{dd4}), we see that
\[
\hat{D}_0(u)\hat{D}_0(v)\hat{F}_{01}= \frac{-1}{p_0(u)-p_0(v)}(\hat{D}_0(u)-\hat{D}_0(v))\hat{F}_{01,01},
\]
and thus
\be
\hat{D}_0(u)\hat{D}_0(v)\hat{D}_0(w)F
=\sum^3_{i=1}res_{p_i}\frac{\hat{D}_0(\la(p))\hat{F}_{01,01}}{(p-p_0(u))(p-p_0(v))(p-p_0(w))}dp.
\label{res1}
\ee
Expanding both side of (\ref{res1}) in the powers of $u$,$v$,$w$,  we get that
\[
\hat{F}_{0i,0j,0k}=\frac{1}{2\pi i}\oint_{C_{\infty}}\frac{\hat{\pa}_{01}\la(p)}{\la^{\prime}(p)}
\hat{\phi}_{0i}\hat{\phi}_{0j}\hat{\phi}_{0k}dp,
\]
where, due to the fact that $0=\hat{\pa}_{01}p(p)=\hat{\pa}_{01}p(\la)+
\pa_\la p(\la)\hat{\pa}_{01}\la$, we have replaced
$\hat{D}_0(\la)\hat{F}_{01,01}=-\hat{\pa}_{01}p(\la)=\hat{\pa}_{01}\la/\la^{\prime}(p)$
in the numerator of (\ref{res1}).
For those cases containing different domain indices, one can derive the formulas by
straightforward calculations. In fact, they can be summarized as a single equation :
\[
\hat{D}_{\alpha}(u)\hat{D}_{\beta}(v)\hat{D}_{\ga}(w)F
=\sum^3_{i=1}res_{p_i}\left(\frac{\hat{D}(\la)\hat{F}_{01,01}}{(p-p_{\alpha}(u))
(p-p_{\beta}(v))(p-p_{\ga}(w))}dp\right),
\]
where $\alpha, \beta, \ga = 0,1,\cdots,N$ and $\hat{D}(\la)$ denotes as a bookkeeping notation
defined by $\hat{D}(\la(p_\alpha))=\hat{D}_\alpha(\la)$.

\subsection{Finite-dimensional reductions}
We now consider finite-dimensional reductions of the Whitham hierarchy.
Suppose $\phi_\alpha(\la)$ are times-independent functions so that
$E(p)=\phi_\alpha(\lambda(p))$ is a meromorphic function of $p$ in $\Omega_\alpha (\alpha=0,\cdots,N)$
with the number of poles being unchanged under the flows $\hat{t}_{\alpha n}$. Then
the sum of the residue in the third derivative of $F$ is now the zeros of $E^\prime(p)$:
\be
\hat{F}_{JKL}=\sum_{E^\prime(p_s)=0}res_{p_s}
\left(\frac{-\hat{\pa}_{01}E(p)}{E^\prime(p)}\hat{\phi}_J(p)\hat{\phi}_K(p)
\hat{\phi}_L(p)dp\right).
\label{resE}
\ee
Such a finite-dimensional reduction imposes a set of additional constraints on the system
and the algebra becomes finite-dimensional.
Since unknowns of the system are functions of the second derivative of the free energy $F$,
one can choose $\hat{F}_{01,P}\equiv \hat{F}_{IP}$ as the independent variables called
primary fields where the time parameters $\{\hat{t}_P\}$ span a finite-dimensional small phase space.
All other $\hat{F}_{ID}$ can be expressed in terms of these primary fields
through a functional relation $Q_D(F_{IP})$ where $D$ denotes those time indices outside
the set $P$. With respect to this separation,
we rewrite the $\hat{F}_{KLM}$ as
\be
\begin{split}
\hat{F}_{KLM}&=\sum_{J\in P}\left(\hat{C}_{KL}^J+
\sum_D\hat{C}^D_{KL}\frac{\pa Q_D}{\pa F_{IJ}}\right)\hat{F}_{JMI}\\
\quad&=\sum_{J\in P} \tilde{\hat{C}}^J_{KL}\hat{F}_{JMI}
\end{split}
\ee
The object $\tilde{\hat{C}}^J_{KL}=\hat{C}_{KL}^J+
\sum_D\hat{C}^D_{KL}\frac{\pa Q_D}{\pa F_{IJ}}$ define the structure constant
of a finite-dimensional
algebra generated by the primary fields. Consequently the finite-dimensional
WDVV equation is now defined by $\tilde{\hat{X}}_{JKLM}\equiv
\sum_{A\in P} \tilde{\hat{C}}^A_{JK}\hat{F}_{ALM}$ $(J,K,L,M\in P)$.
Following \cite{Boy01}, one can show that
this $\tilde{\hat{X}}_{JKLM}$ is totally symmetric with respect to
all the indices in the set $P$ by proving that
$\tilde{\hat{X}}_{JKLM}=\hat{X}_{JKLM}$ for $J,K,L,M\in P$.
This completes the verification for the associativity equations corresponding to
finite-dimensional reductions of the dHirota equations.
Define a scalar product :
\be
\langle f\cdot g\rangle = \sum_{E'(p_s)=0} res_{p_s} \left(\frac{-\hat{\pa}_{01}E(p)}{E^\prime(p)}f(p)g(p)dp\right),
\ee
then we see that $\eta_{JK}=\hat{F}_{JKI}=
\langle\hat{\phi}_J\cdot\hat{\phi}_K\rangle$
because of $\hat{\phi}_{I}=1$. Therefore
\be
\begin{split}
\hat{F}_{JKL}&=\langle \hat{\phi}_J\hat{\phi}_K\cdot\hat{\phi}_L\rangle
=\sum_{A\in P} \tilde{\hat{C}}^A_{JK}\langle \hat{\phi}_A \cdot\hat{\phi}_L\rangle\\
&=\sum_{A\in P}\tilde{ \hat{C}}^A_{JK}\hat{F}_{ALI}.
\end{split}
\ee
This gives another proof of the WDVV equation since now it reads
$\tilde{\hat{X}}_{JKLM}=\langle\hat{\phi}_J\hat{\phi}_K\cdot
\hat{\phi}_L\hat{\phi}_M\rangle$ and the symmetry
with respect to the permutation is now obvious.
Let us mention the well-known finite-dimensional reduction constructed by Krichever\cite{Kr94},
in which $\phi_\alpha(\la)=\la_\alpha^{n_\alpha}$ and $E(p)$
is a rational function of the form
\be
E(p)=p^{n_0}+u_{n_0 -2}p^{n_0 -2}+\cdots+u_{0}+
\sum^N_{\alpha =1}\sum^{n_{\alpha}}_{s=1}\frac{v_{\alpha s}}{(p-q_\alpha)^s}
\ee
which provides the algebraic orbits(or Lax function) of genus-zero Whitham hierarchy.
Note that the number of zeros $p_s$ in (\ref{resE}) is equal to the number of unknowns
$\{u_{n_0},\cdots,u_{n_0 -2},v_{\alpha s},q_\alpha\}$ and the small phase space
is characterized by the time-parameters $\hat{t}_P$ with $P$ being the set
$\{(0,i),i=1,\cdots,n_0 -1;(\alpha,m),\alpha = 1,\cdots,N;m=0,\cdots, n_\alpha\}$.
\section{ Concluding Remarks}
We have demonstrated that the kernel formula approach developed by Carroll and Kodama for the dKP hierarchy
can be generalized well to the universal Whitham hierarchy.
The obtained dHirota equations are exactly the same as those derived by Takasaki and Takebe
using the notion of Faber polynomials.
We extract the structure constants of an associate algebra from the dHirota equations and
verify the associativity equations of the algebra.
Finally, we give a realization of the associative algebra with the structure constants
expressed in terms of the residue formula and verify the associativity equations for
finite-dimensional reductions.
We like to emphasize that the kernel formula approach indeed provides an efficient way
to derive the dHirota equations of the genus-zero Whitham hierarchy.
It would be interesting to generalize and/or
reformulate the kernel formula approach to higher genus \cite{Kr94}. We hope to address this
issue in the future work.

{\bf Acknowledgments\/}\\
This work is supported by the National Science Council of Taiwan
under Grant  NSC97-2112-M-194-002-MY3.

\end{document}